\documentclass[12pt,preprint]{aastex}%
\usepackage{amsfonts}
\usepackage{amssymb}
\usepackage{graphicx}
\usepackage{amsmath}%
\slugcomment{Preprint}
\shorttitle{Gravity-powered Jet Dynamo}
\shortauthors{Bellan}

\begin{document}
%
\makeatletter\def\stackunder#1#2{\mathrel{\mathop{#2}\limits_{#1}}}
\makeatother
%

\title
{Enrichment of the dust-to-gas mass ratio in  Bondi/Jeans accretion/cloud systems due to unequal changes in dust and gas incoming velocities}%
%

\author{P. M. Bellan}%
%

\affil{Applied Physics, Caltech, Pasadena CA 91125, USA}%
%

\email{pbellan@caltech.edu}%
%

\begin{abstract}
The ratio of the Bondi and Jeans lengths is used to develop a cloud-accretion
model that describes both an inner Bondi-type regime where gas pressure is
balanced by the gravity of a central star and an outer Jeans-type regime where
gas pressure is balanced by gas self-gravity. The gas density profile provided
by this model makes a smooth transition from a wind-type inner solution to a
Bonnor-Ebert type outer solution. It is shown that high-velocity dust
impinging on this cloud will tend to pile-up due to having a different
velocity profile than gas so that the dust-to-gas ratio is substantially
enriched above the 1\% ISM\ level.
\end{abstract}
\ \ \ \ \ \ \ \ \ %


\section{Introduction}

Laboratory-scale plasma jets have been produced by a magnetohydrodynamic
mechanism believed analogous to the mechanism responsible for driving
astrophysical jets
\citep{Hsu2002,Bellan2005}%
. The laboratory jets are driven by capacitor bank power supplies that provide
poloidal and toroidal magnetic fields and the jet acceleration mechanism can
be considered as due to the pressure of the toroidal magnetic field inflating
flux surfaces associated with the poloidal magnetic field. If these jets are
indeed related to astrophysical jets, the obvious question arises as to what
constitutes the power supply responsible for the toroidal and poloidal
magnetic fields in the astrophysical situation. Existing models of
astrophysical jets typically assume the poloidal field is simply given and
that the toroidal field results from the rotation of an accretion disk
twisting up the assumed poloidal field. The author has developed an alternate
model which postulates that the toroidal and poloidal field result instead
from a dusty plasma dynamo mechanism that converts the gravitational energy of
infalling dust grains into an electrical power source that drives poloidal and
toroidal electric currents creating the respective toroidal and poloidal
fields. A brief outline of how infalling charged dust can drive poloidal
currents has been presented in
\citet{Bellan2007}%
.

An important requirement of the model is that there should be sufficient
infalling dust to provide the jet power. It has been well established that the
dust-to-gas mass ratio in the Interstellar Medium (ISM) is 1\%. If one
assumes, as has been traditional, that this ratio holds throughout the
accretion process, then the gravitational energy available from infalling dust
would be insufficient. However,
\citet{Fukue2001}
has recently shown via numerical solution of coupled dust and gas equations of
motion that the dust-to-gas ratio can become substantially enriched during
Bondi-type accretion.

Star formation has also been previously examined from a molecular cloud
physics point of view which differs significantly from the Bondi accretion
point of view in the treatment of self-gravity and inflows. Molecular cloud
physics is clearly important because observations indicate that stars form
within the dense cores of molecular clouds (e.g., see
\citet{Evans2001}%
). Analysis of the force balance in molecular clouds shows that the cloud
radial density profile can be characterized by the Bonnor-Ebert sphere
solution
\citep{Ebert1955,Bonnor1956}%
. This solution does not take into account flows associated with accretion.
Dust dynamics is also not taken into account; instead it is typically assumed
that the dust-to-gas mass ratio is fixed at 1\% for all radii.

To summarize the above discussion, we note that the Bondi-type analysis
reported by
\citet{Fukue2001}
emphasizes inflow physics and dust-gas coupling but does not take into account
self-gravity whereas molecular cloud analysis such as used by
\citet{Evans2001}
emphasizes self-gravity, but does not take into account inflow or dust-gas coupling.

This paper will address the physics of coupled dust and gas accretion using a
methodology similar in concept to that presented by
\citet{Fukue2001}%
, but extended to bridge the gap between Bondi accretion models and molecular
cloud force balance models. \ The enrichment mechanism observed by
\citet{Fukue2001}
will be examined in detail and will be shown to result from the inherently
non-uniform nature of the Bondi/molecular cloud/ISM system. Properties of dust
and gas for radii ranging from the cloud-ISM interface to the inner Bondi
region will be considered by examining a sequence of successively smaller
concentric regions. The basic character of each region and its scale, defined
in terms of the nominal radial distance from a central object star, is as follows:

\begin{description}
\item \textit{ISM scale:} The outermost scale is that of the Interstellar
Medium (ISM). The ISM\ has a gas density $\sim$10$^{7}$ m$^{-3}$, a
dust-to-gas mass ratio of 1 percent, a gas temperature $T_{g}^{ISM}\sim100$
K$,$ and is optically thin. The ISM is assumed to be spatially uniform and to
bound a molecular cloud having radius $r_{edge}$.

\item \textit{Molecular cloud scale:}\ The molecular cloud scale has much
higher density than the ISM and is characterized by force balance between gas
self-gravity\ and gas pressure. The molecular cloud scale is sub-divided into
a large, radially non-uniform low-density outer region and a small,
approximately uniform, high-density inner core region. Clouds have a
characteristic scale given by the Jeans length $r_{J}.$ The radial dependence
of density is provided by the Bonnor-Ebert sphere solution which acts as the
outer boundary of the Bondi accretion scale.

\item \textit{Bondi accretion scale:} The Bondi accretion scale
\citep{Bondi1952}
is $\sim r_{B}$ which is sufficiently small that gas self-gravity no longer
matters so equilibrium is instead obtained by force balance between gas
pressure and the gravity of a central object assumed to be a star having mass
$M_{\ast}$. The Bondi scale is sub-divided into three concentric radial
regions: an outermost region where the gas flow is subsonic, a critical
transition radius at exactly $r_{B}$ where the flow is sonic, and an innermost
region where the gas flow is free-falling and supersonic.

\item \textit{Collisionless dusty plasma scale (to be considered in a future
publication): } Free-falling dust grains collide with each other in one of the
above scales and coagulate to form large-radius grains which are collisionless
and optically thin. The optically thin dust absorbs UV photons from the star,
photo-emits electrons and becomes electrically charged. The charged dust
grains are subject to electromagnetic forces in addition to gravity. Motions
of charged dust grains relative to electrons result in electric currents with
associated poloidal and toroidal magnetic fields [see preliminary discussion
in
\citet{Bellan2007}%
]. \ \ 

\item \textit{Jet scale (to be considered in a future publication): }The
electric currents interact with the magnetic fields to produce
magnetohydrodynamic forces which drive astrophysical jets in a manner
analogous to that reported in \
\citet{Hsu2002}
and
\citet{Bellan2005}%
.
\end{description}

\bigskip The separation-of-scales requirement $r_{J}>>r_{B}$ implies existence
of a small parameter%
\begin{equation}
\varepsilon_{BJ}\ =\frac{r_{B}}{r_{J}}\ \label{eps}%
\end{equation}
quantifying the separation between the Bondi and Jeans scales. For purposes of
relating the Bondi and Jeans scales to each other it is convenient to
introduce a geometric-mean scale with characteristic length given by
\begin{equation}
r_{gm}=\sqrt{r_{B}r_{J}}=\sqrt{\varepsilon_{BJ}}r_{J}\ \label{def rgm}%
\end{equation}
in which case%
\begin{equation}
r_{B}\ll r_{gm}\ll r_{J}. \label{separation}%
\end{equation}

Gas and dust must be considered separately in each scale. We assume that gas
motion influences dust motion but not vice-versa so that the dust dynamics can
be ascertained after gas dynamics has been determined. This assumption is
appropriate so long as dust mass and energy densities are small compared to
corresponding gas densities or if the dust is decoupled from the gas. The
sequence \ of scales is summarized in Tables \ref{gas table} and
\ref{dust table}. The numerical values of table elements with asterisks will
be predicted by the model to be presented here while table elements with
filled-in numbers represent\ prescribed physical boundary conditions. Here
$m_{g}$ is the gas molecular mass, $n_{g}$ is the gas density, $T_{g}$ is the
gas temperature, $u_{g,d}$ are the radially inward gas and dust fluid
velocities, and $\ $ $\rho_{g,d}$ are the gas and dust mass densities.
Quantities with superscripts `ISM' are evaluated in the ISM, un-superscripted
quantities refer to Bondi or molecular cloud regions.

Our methodology will be conceptually similar to
\citet{Fukue2001}
but will also differ in important ways.
\citet{Fukue2001}
used a wind equation scheme to model Bondi-type physics and, unlike the
analysis to be presented here, assumed the Bondi accretion region was directly
bounded by the ISM, i.e., no molecular cloud region with Jeans-type scaling
was taken into account. In addition, Fukue assumed that (i) the gas was heated
by the combined effects of adiabatic compression and friction due to dust-gas
collisions, and (ii) the entire region was optically thin so that the dust was
subject to radiation pressure. Because of the complexity introduced by the
heating of the gas, Fukue obtained results via numerical solution of three
coupled differential equations (gas momentum, dust momentum, gas heating). Our
approach will differ by assuming that (i) the gas is isothermal, (ii) a
molecular cloud region lies between the Bondi accretion region and the ISM,
and (iii) the system is optically thick outside the inner part of the Bondi
region so that dust is shielded from optical radiation until it penetrates to
the inner part of the Bondi region. Furthermore, rather than using numerical
solutions, we will attempt analytic solutions as much as possible, a goal made
feasible by the isothermal assumption. We believe our assumptions (i)-(iii)
reasonably correspond to observations that gas is nearly isothermal and that
stars form inside the cores of optically thick molecular clouds.

\bigskip%
\begin{table}[tbp] \centering
\begin{tabular}
[c]{lllll}\hline\hline
region & location & $n_{g}$ & $T_{g}$ & $u_{g}$\\\hline\hline
ISM & $r>r_{edge}$ & 10$^{7}$ m$^{-3}$ & 100 K & *\\
B-E sphere & $r_{gm}<r<r_{edge}$ & * & 10 K & *\\
Bondi-Jeans interface & $r_{gm}=\sqrt{r_{B}r_{J}}$ & * & \multicolumn{1}{c}{"}
& *\\
Bondi subsonic & $r_{B}<r<r_{gm}$ & * & \multicolumn{1}{c}{"} & $<\sqrt{\kappa
T_{g}/m_{g}}$\\
Bondi transonic & $r=r_{B}$ & * & \multicolumn{1}{c}{"} & $\sqrt{\kappa
T_{g}/m_{g}}$\\
Bondi supersonic & $r<r_{B}$ & * & \multicolumn{1}{c}{"} & $>\sqrt{\kappa
T_{g}/m_{g}}$%
\end{tabular}
\caption{Sequence of regions for gas, * indicates quantity to be discussed/calculated in text}\label{gas table}%
\end{table}%
%

\begin{table}[tbp] \centering
\begin{tabular}
[c]{llll}\hline\hline
region & location & $\rho_{d}/\rho_{g}$ & $u_{d}$\\\hline\hline
ISM & $r>r_{edge}$ & 0.01 & 3 km/s (turbulent acceleration)\\
B-E sphere & $\sqrt{r_{B}r_{J}}<r<r_{edge}$ & * & slowing down by gas\\
Bondi-Jeans interface & $r_{gm}=\sqrt{r_{B}r_{J}}$ & * & entrained with gas\\
Bondi subsonic & $r_{B}<r<r_{gm}$ & * & \multicolumn{1}{c}{"}\\
Bondi transonic & $r=r_{B}$ & * & \multicolumn{1}{c}{"}\\
Bondi supersonic & $r<r_{B}$ & * & \multicolumn{1}{c}{"}%
\end{tabular}
\caption{Sequence of regions for dust physics, * indicates to be discussed/calculated in text}\label{dust table}%
\end{table}%

\section{Gas and dust equations}

The steady-state equation of motion for spherically symmetric gas is
\begin{equation}
\rho_{g}u_{g}\frac{du_{g}}{dr}\mathbf{\ }=-\frac{\partial P_{g}}{\partial
r}\ -\rho_{g}\frac{MG}{r^{2}}\ +f_{drag}. \label{gas motion}%
\end{equation}
Here $u_{g}$ is the radial fluid velocity of the gas, $P_{g}=n_{g}\kappa
T_{g}$ is the gas pressure, $\rho_{g}$ is the gas mass density,
\begin{equation}
M=\ M_{\ast}+\int_{r_{\ast}}^{r}4\pi\xi^{2}\left(  \rho_{g}(\xi)+\rho_{d}%
(\xi)\right)  d\xi\ \label{M}%
\end{equation}
is the mass inside a sphere of radius $r,$ $\ \rho_{d}\ $is the dust mass
density, $M_{\ast}$ and $r_{\ast}$ are the mass and radius of the central
object star, and $f_{drag}$ is the drag force on gas due to collisions with
dust.
\citet{Lamers1999}
give the drag force to be%
\begin{equation}
f_{drag}=-(u_{g}-u_{d})\frac{\rho_{g}\rho_{d}}{m_{d}}\sigma_{d}\sqrt{c_{g}%
^{2}+(u_{d}-u_{g})^{2}}. \label{fdrag}%
\end{equation}
The dust behaves as a zero-pressure fluid, so its steady-state equation of
motion is%
\begin{equation}
\rho_{d}u_{d}\frac{du_{d}}{dr}\mathbf{\ }=\ \ -\rho_{d}\frac{MG}{r^{2}%
}\ -f_{drag}. \label{dust motion}%
\end{equation}
In the limit that the integral in Eq.\ref{M} can be ignored, these are the
same momentum equations considered by
\citet{Fukue2001}%
.

Observations indicate that the gas temperature of clouds, cores, and accretion
disks lies in the range 10-30 K indicating that the gas can be considered
approximately isothermal. If the gas were adiabatic, i.e., if $P_{g}\sim
\rho_{g}^{5/3}$ then the gas temperature would vary as $T_{g}\sim n^{2/3}.$
Since there is at least a $10^{3}$ increase in gas density from the cloud edge
to the Bondi region inner free-fall region, an adiabatically compressed gas
volume element would have its temperature increase at least one-hundred fold
as it moved inwards. If in addition, there were heating of gas due to dust
frictional drag, the gas temperature would increase even more. For example,
the sound speed in Fukue's Fig. 1 increases by a factor of approximately 40
from right to left corresponding to a factor of 1600 increase in temperature.
Thus Fukue's Fig. 1 analysis predicts that the gas temperature increases from
100 K to $1.6\times10^{5}$ K a temperature at which the molecular hydrogen
would not only have become disassociated but would have turned into plasma.
Since the observed gas temperature is not\ $1.6\times10^{5}$ K but in fact is
clamped at 10-30 K, any heat energy resulting from adiabatic compression or
friction must be immediately lost, presumably via molecular line emission. In
order to conform to observations we therefore assume the gas is isothermal, an
approximation which has the incidental, yet fortuitous side effect of making
an analytic approach to the problem feasible.

Since we are assuming that the dust is a perturbation on the gas, we first
solve the gas equation ignoring the dust, and then use gas equation solutions
as inputs for the dust equation. This approach is self-consistent so long as
the dust is approximately decoupled from the gas or if the dust mass and
energy densities are much less than the gas mass and energy densities. The
approach fails in situations where the dust is coupled to the gas and attains
mass or energy densities comparable to the gas. However, the point at which
failure occurs is interesting and can be considered a useful prediction of the model.

We now consider gas dynamics while ignoring the dust in which case the gas
equation of motion reduces to
\begin{equation}
\rho_{g}u_{g}\frac{du_{g}}{dr}\mathbf{\ }=-\ c_{g}^{2}\frac{d\rho_{g}%
}{\partial r}\ -\frac{\ \rho_{g}G}{r^{2}}\left(  M_{\ast}+\int_{r_{\ast}}%
^{r}4\pi\xi^{2}\rho_{g}(\xi)d\xi\right)  \ \label{gas motion 2}%
\end{equation}
where%
\begin{equation}
c_{g}=\sqrt{\frac{\kappa T_{g}}{m_{g}}\ } \label{cg}%
\end{equation}
is the gas thermal velocity.

Equation \ref{gas motion 2} has two regimes of interest, namely

\begin{enumerate}
\item an inner regime where $r$ is so small that $M_{\ast}\gg\int_{r_{\ast}%
}^{r}4\pi\xi^{2}\rho_{g}(\xi)d\xi$ and

\item an outer regime where $r$ is so large that $M_{\ast}\ll\int_{r_{\ast}%
}^{r}4\pi\xi^{2}\rho_{g}(\xi)d\xi.$
\end{enumerate}

The former regime leads to a Bondi accretion situation characterized by the
wind equation discussed by
\citet{Fukue2001}
while the latter leads to a Jeans-scale problem with Bonnor-Ebert spheres as
the solution. We will examine these regimes separately and then connect them
using an asymptotic technique.

Because the configuration is assumed to be steady-state, the\ gas and dust
equations of continuity give the respective accretion rates%
\begin{equation}
\dot{M}_{g}=-4\pi r^{2}\rho_{g}u_{g}=const. \label{Mdot gas}%
\end{equation}%
\begin{equation}
\dot{M}_{d}=-4\pi r^{2}\rho_{d}u_{d}=const. \label{Mdot dust}%
\end{equation}
where $u_{g}$ and $u_{d}$ are negative, corresponding to radial inward motion.

\section{Gas}

\subsection{Bondi accretion region \label{Bondi regime}}

In this region it is convenient to normalize the velocity to $c_{g}$ and
lengths to the Bondi radius\ $r_{B}$, defined as
\begin{equation}
r_{B}=\frac{M_{\ast}G}{2c_{g}^{2}}\ . \label{rB}%
\end{equation}
A bar will denote normalized quantities and to avoid confusing minus signs,
the normalized velocity is defined to be positive so%
\begin{equation}
\bar{u}_{g}=\left\vert u_{g}\right\vert /c_{g} \label{u norm}%
\end{equation}%
\begin{equation}
\bar{r}=r/r_{B}. \label{r norm}%
\end{equation}
Also the gas mass density is normalized to its value at $r_{gm}$, the
geometric mean of the Bondi and Jeans scales (see Eq.\ref{def rgm}), i.e.,
\begin{equation}
\bar{\rho}_{g}=\rho_{g}/\rho_{gm}. \label{rho norm}%
\end{equation}
Thus $\rho_{gm}$ is the mass density at a radius much larger than the Bondi
length, but much smaller than the Jeans length (see Eq.\ref{separation}). The
normalized form of Eq.\ref{gas motion 2} in the Bondi accretion region is thus%
\begin{equation}
\bar{u}_{g}\frac{d\bar{u}_{g}}{d\bar{r}}\mathbf{\ }=-\ \frac{d\ln\bar{\rho
}_{g}}{d\bar{r}}\ -\frac{2}{\ \bar{r}^{2}}. \label{Bondi norm motion}%
\end{equation}
The derivative of Eq.\ref{Mdot gas} gives
\begin{equation}
\ \ \frac{d\ln\bar{\rho}_{g}\ }{d\bar{r}}=-\frac{1}{\bar{u}_{g}}\frac{d\bar
{u}_{g}}{d\bar{r}}-\frac{2}{\bar{r}}\ \ . \label{derivative Mdot}%
\end{equation}
Combining Eqs.\ref{Bondi norm motion} and \ref{derivative Mdot} gives the
isothermal wind equation
\citep{Lamers1999}
\begin{equation}
\left(  \bar{u}_{g}^{2}-1\right)  \ \frac{d\bar{u}_{g}}{d\bar{r}}%
\mathbf{\ }=\ 2\bar{u}_{g}\left(  \frac{1}{\bar{r}}\ -\frac{1}{\bar{r}^{2}%
}\right)  . \label{wind}%
\end{equation}
Equation \ref{wind} gives the condition that $\bar{u}_{g}=1$ must occur when
$\bar{r}=1$ in order for $d\bar{u}_{g}/d\bar{r}$ to be non-singular at
$\bar{r}=1.$ \ Equation \ref{wind} can be directly integrated to give%
\begin{equation}
\frac{\bar{u}_{g}^{2}}{2}-\ln\bar{u}_{g}=2\ln\bar{r}+\frac{2}{\bar{r}}+A
\label{integrate wind}%
\end{equation}
where $A$ is a constant to be determined. The sonic condition $\bar{u}_{g}=1$
at $\bar{r}=1\,$\ gives $A=-3/2$ $\ $so Eq.\ref{integrate wind} becomes%
\begin{equation}
\bar{u}_{g}\bar{r}^{2}-\exp\left(  \frac{\bar{u}_{g}^{2}}{2}-\frac{2}{\bar{r}%
}+\frac{3}{2}\right)  =0. \label{solve wind}%
\end{equation}
Equation \ref{solve wind}, the solution to Eq.\ref{wind}, is a transcendental
expression relating $\bar{u}_{g}$ and $\bar{r}$ and having two distinct roots.
One root has $\bar{u}_{g}$ a monotonically increasing function of $\bar{r}$
and one root has $\bar{u}_{g}$ a monotonically decreasing function. Both roots
have $\ \bar{u}_{g}=1$ at $\bar{r}=1\ .$ The root where $\bar{u}_{g}$ is a
monotonically increasing function of $\bar{r}$ (subsonic at small $\bar{r},$
supersonic at large $\bar{r}$) is relevant to stellar winds, while the
monotonically decreasing root (supersonic at small $\bar{r}$, subsonic at
large $\bar{r}$) is relevant to the accretion problem discussed here and by
Fukue. Since $\bar{u}_{g}$ is subsonic at large $\bar{r}$ and supersonic at
small $\bar{r},$ Eq.\ref{integrate wind} has the following limiting forms%
\begin{equation}
\ln\bar{u}_{g}\ =\ln2-\frac{1}{2}\ln\bar{r}\ \ \mathrm{\ for\quad}\bar{r}%
\ll1\ \label{log dependence small r}%
\end{equation}%
\begin{equation}
\ln\bar{u}_{g}=-\ 2\ln\bar{r}\ +\frac{3}{2}\ \ \mathrm{for\quad}\bar{r}\gg1.
\label{log dependence large  r}%
\end{equation}
The change in the slope of $\ \ln\bar{u}_{g}\ $\ as a function of $\ln\bar{r}$
from $-1/2$ at small $\bar{r}$ to $-2$ at large $\bar{r}$ is evident in Fig. 1
of Fukue.

The mass accretion rate can be determined by evaluation at $\bar{r}\gg1$ since
in this limit Eq.\ref{solve wind} gives%
\begin{equation}
\bar{u}_{g}\ =\ \ \frac{1}{\bar{r}^{2}}\exp\left(  -\frac{2}{\bar{r}}+\frac
{3}{2}\right)  \ \ \ \mathrm{for\quad}\bar{r}\gg1. \label{ug large r}%
\end{equation}
Using Eq.\ref{ug large r} in Eq.\ref{Mdot gas} and then evaluating
at$\ r_{gm}\ $ gives the Bondi mass accretion rate
\begin{equation}
\dot{M}_{g}=\ \pi\frac{\eta\rho_{gm}M_{\ast}^{2}G^{2}}{c_{g}^{3}%
}\ \label{Bondi accretion rate basic}%
\end{equation}
where
\begin{equation}
\eta(\varepsilon_{BJ})=\exp\left(  -2\sqrt{\varepsilon_{BJ}}+\frac{3}%
{2}\right)  \ \label{define eta}%
\end{equation}
is a dimensionless quantity of order unity which occurs repeatedly in the
analysis to follow. The radial dependence of the gas mass density in the Bondi
regime is given by
\begin{equation}
\rho_{g}(r)=\ \frac{\eta\rho_{gm}M_{\ast}^{2}G^{2}}{4r^{2}\left\vert
u_{g}(r)\right\vert c_{g}^{3}}\ . \label{ng}%
\end{equation}
$\rho_{g}$ becomes independent of $r$ in the large $r$ limit of the Bondi
region since in this limit Eq.\ref{ug large r} shows that $\bar{u}_{g}%
\sim1/\bar{r}^{2}$. This independence permits matching the large $r\ $ limit
of the Bondi region solution to the small $r\ $ limit of the molecular cloud
region solution (given in Sec.\ref{Jeans regime} below) since both these
solutions are independent of $r.$

In the small $\bar{r}$ limit, Eq. \ref{integrate wind} gives
\begin{equation}
\bar{u}_{g}\ =\ 2/\bar{r}^{1/2}\ \label{free-fall}%
\end{equation}
which corresponds to the free-fall velocity. Combining this with
Eqs.\ref{Mdot gas} and \ref{Bondi accretion rate basic} gives the mass density
for $r<<r_{B}$ to be
\begin{equation}
\ \bar{\rho}_{g}(\bar{r})=\frac{\eta}{2\bar{r}^{3/2}}.
\label{inner mass density}%
\end{equation}
The enclosed gas mass at $r_{B}$ is
\begin{equation}
\int_{0}^{r_{B}}4\pi r^{2}\rho_{g}dr\simeq4\pi\rho_{gm}r_{B}^{3}\int_{0}%
^{1}\bar{r}^{2}\bar{\rho}_{g}d\bar{r}=\frac{4}{3}\pi\eta\rho_{gm}r_{B}%
^{3}\ .\ \label{Mass enclosed by rB}%
\end{equation}

\subsection{Molecular cloud region\label{Jeans regime}}

Again ignoring dust mass and dust drag, but now assuming $\bar{r}$ is
sufficiently large that $M_{\ast}\ll\int_{r_{\ast}}^{r}4\pi\xi^{2}\left(
\rho_{g}(\xi)\right)  d\xi$, Eq.\ref{gas motion 2} becomes%

\begin{equation}
\ \frac{c_{g}^{2}}{\rho_{g}}\ \frac{d\rho_{g}}{dr}\ +\frac{G}{r^{2}}%
\ \int_{r_{\ast}}^{r}\ 4\pi\xi^{2}\rho_{g}(\xi)d\xi\ \ =0\ \label{B-E start}%
\end{equation}
where the subsonic condition $u_{g}^{2}<<c_{g}^{2}$ has been used. Gas
pressure is now balanced by self-gravity rather than by the central object
gravity. Equation \ref{B-E start} can be recast in standard form as
\begin{equation}
\frac{c_{g}^{2}}{4\pi G}\frac{1}{r^{2}}\ \frac{d}{dr}\left(  \frac{r^{2}}%
{\rho_{g}}\frac{d\rho_{g}}{dr}\right)  \ +\rho_{g}\ =0. \label{B-E 2}%
\end{equation}
We now define a new dimensionless length suitable for Eq.\ref{B-E 2}, namely%
\begin{equation}
\tilde{r}=\frac{r}{r_{J}} \label{r tilde}%
\end{equation}
where
\begin{equation}
r_{J}=\frac{c_{g}}{\sqrt{4\pi\rho_{gm}G}} \label{Jeans radius}%
\end{equation}
is the Jeans length.

Equation \ref{B-E 2} assumes the dimensionless form
\begin{equation}
\ \frac{1}{\ \tilde{r}^{2}}\frac{d}{d\tilde{r}}\left(  \frac{\tilde{r}^{2}%
}{\bar{\rho}_{g}}\frac{d\bar{\rho}_{g}}{d\tilde{r}}\right)  \ +\bar{\rho}%
_{g}\ =0. \label{B-E 3}%
\end{equation}
For small $\bar{r}$, we wish to have a solution which is approximately
constant in order to be consistent with observations. This constant should be
unity in order to connect with the outer limit of the Bondi solution since the
outer limit of the Bondi solution has been constructed so as to give
$\rho\rightarrow\rho_{gm}$ for $r=r_{gm}.$ An approximate solution to
Eq.\ref{B-E 3} that does this is $\bar{\rho}_{g}=1/(1+\tilde{r}^{2}/6)$ but
this solution fails at large $\bar{r}.$ However, Equation \ref{B-E 3} has the
exact solution $\bar{\rho}_{g}=2/\bar{r}^{2}$ which is valid at all $\bar{r}.$
Thus, a solution which behaves as $\bar{\rho}_{g}=1/(1+\tilde{r}^{2}/6)$ for
small $\tilde{r}$ and as $\bar{\rho}_{g}=2/\tilde{r}^{2}$ for large $\tilde
{r}$ would do the job provided these sub-solutions smoothly merge into each
other and do not introduce any spurious effects at intermediate values of
$\tilde{r}.$ The Pad\'{e} approximation
\begin{equation}
\bar{\rho}_{g}(\bar{r})=\frac{1}{1+\frac{\tilde{r}^{2}}{6\ }\left(  1+2\left(
\frac{\tilde{r}^{2}}{15\ +\tilde{r}^{2}\ }\right)  ^{2}\right)  }
\label{Pade solution}%
\end{equation}
agrees with a direct numerical solution of Eq.\ref{B-E 3} with an error less
than 20\% in the range $0<\tilde{r}<9,\ $\ an accuracy which is more than
adequate for the present analysis. The factor $\left(  \tilde{r}%
^{2}/(15+\tilde{r}^{2})\right)  ^{2}$ can be thought of as a `switch' which
causes $\tilde{r}^{2}/6$ to become $\tilde{r}^{2}/2$ at large $\tilde{r}^{2}.$

At $r_{gm},$ the value of $\tilde{r}$ is $\tilde{r}=r_{gm}/r_{J}%
=\sqrt{\varepsilon_{BJ}}$ and so Eq.\ref{B-E 3} matches to the Bondi solution
at $r_{gm}$ since by assumption $\varepsilon_{BJ}\ll1.$

\begin{figure}
[ptb]
\begin{center}
\includegraphics[
height=7.9208in,
width=5.7173in
]%
{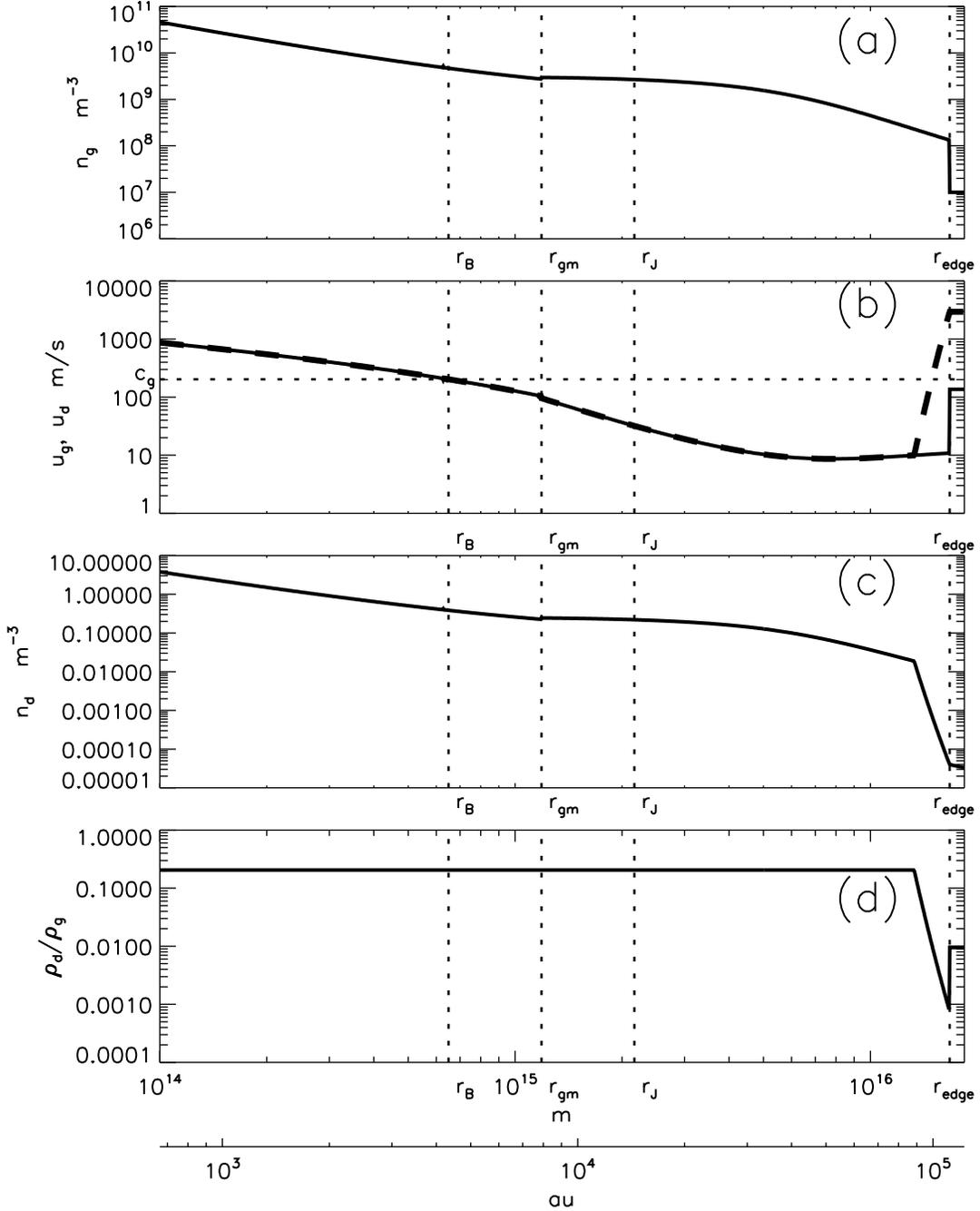}%
\caption{Model predictions assuming $\ $ $\varepsilon_{BJ}=0.3,$ $M_{\ast
}=0.4M_{\odot},$ $T_{g}=10$ K, $T_{g}^{ISM}=100$ K, and $n_{g}^{ISM}=10^{7}%
$m$^{-3}.$ (a) gas density, (b) solid line gives gas velocity, dashed line
gives dust velocity, (c) dust density, (d) dust-to-gas mass ratio.}%
\label{f1.eps}%
\end{center}
\end{figure}

The combined solution spanning the range $0<r<9r_{J}$ is therefore \ \ \ \ \
\begin{equation}
\rho_{g}(r)=\left\{
\begin{array}
[c]{ll}%
\ \frac{\eta M_{\ast}^{2}G^{2}}{4r^{2}\left\vert u_{g}(r)\right\vert c_{g}%
^{3}}\rho_{gm} & \mathrm{for\quad}r<r_{gm}\\
& \\
\frac{\rho_{gm}}{1+\frac{r^{2}}{6r_{J}^{2}}\left(  1+2\left(  \frac{r^{2}%
}{15r_{J}^{2}+r^{2}\ }\right)  ^{2}\right)  }\  & \mathrm{for\quad}%
r_{gm}<r<9r_{J}%
\end{array}
\right.  \ \label{solve rhog}%
\end{equation}
where $u_{g}(r)$ is given by Eq.\ref{solve wind} and both outer and inner
solutions converge to $\rho_{gm}$ at $r=r_{gm}.$ Figure \ref{f1.eps}(a) plots
$n_{g}(r)=\rho_{g}(r)/m_{g}$ using Eq.\ref{solve rhog} with the assumptions
that $\varepsilon_{BJ}=0.3,$ $M_{\ast}=0.4M_{\odot},$ $T_{g}=10$ K,
$T_{g}^{ISM}=100$ K, and $n_{g}^{ISM}=10^{7}$m$^{-3}.$ The values of $r_{B},$
$r_{gm},$ and $r_{J}$ are indicated by vertical dashed lines. The solid line
in Fig. \ref{f1.eps}(b) plots the corresponding gas velocity $u_{g}(r);$ note
that $u_{g}=c_{g}\ $ at $r_{B}$ and that the gas is subsonic to the right of
$r_{B}$.%

Specification of $\varepsilon_{BJ}\ $ determines $\rho_{gm}$ since
substitution of Eqs.\ref{rB} and \ref{Jeans radius} into Eq.\ref{eps} gives
\begin{equation}
\ \rho_{gm}=\varepsilon_{BJ}^{2}\frac{c_{g}^{6}}{\pi M_{\ast}^{2}G^{3}}.
\label{density limit}%
\end{equation}
The separation-of-scales condition is consistent with the assumption
$\ M_{\ast}\ll\int_{r_{\ast}}^{r}4\pi\xi^{2}\rho_{g}(\xi)d\xi$ \ for
$\ r\gtrsim r_{J}$ since $\ \int_{r_{\ast}}^{r_{J}}4\pi\xi^{2}\rho_{g}%
(\xi)d\xi\simeq4\pi\rho_{gm}r_{J}^{3}/3=M_{\ast}/6\varepsilon_{BJ}.$

Equation \ref{density limit} can be used with Eq.\ref{Mass enclosed by rB} to
calculate the total gas mass enclosed at $r_{B}$ to be
\begin{equation}
\int_{0}^{r_{B}}4\pi r^{2}\rho_{g}dr=\ \frac{\eta}{6}\varepsilon_{BJ}%
^{2}M_{\ast} \label{enclosed mass at rB 2}%
\end{equation}
which is consistent with the Bondi region assumption that $\int_{0}^{r_{B}%
}4\pi r^{2}\rho_{g}dr\ll M_{\ast}.$

Equations \ref{Bondi accretion rate basic}\ and \ref{density limit} show that
the gas mass accretion rate can be expressed as
\begin{equation}
\dot{M}_{g}=\ \frac{c_{g}^{3}}{G}\varepsilon_{BJ}^{2}\eta. \label{solve Mdot}%
\end{equation}
The gas fluid velocity at $r_{gm}$ is
\begin{equation}
u_{g\ }(r_{gm})=-\frac{\dot{M}_{g}}{4\pi r_{gm}^{2}\rho_{gm}}=-\varepsilon
_{BJ}\eta c_{g}. \label{ug at rgm}%
\end{equation}

Bondi accretion will presumably increase $M_{\ast}.$ Because of the existence
of jets, not all accreting matter will do this. However, for purposes of
estimation, if one assumes that all accreting material causes an increase in
$M_{\ast}$, then $r_{B}$ which is proportional to $M_{\ast}$ will increase. On
the other hand $r_{J}$ does not depend on $M_{\ast}$ and so will remain
constant. Thus, $r_{B}/r_{J}$ will be a slowly increasing function of time and
eventually the presumption that\ $r_{B}\ll r_{J}$ fails.

\bigskip

\subsection{Interface between molecular cloud region and the ISM}

Although gas density and temperature in reality change gradually on entering
the cloud from the ISM, for simplicity we assume here\ that these changes
occur in a narrow layer. Integration of Eq.\ref{gas motion} across this layer
while taking into account mass conservation gives continuity of $\rho_{g}%
u_{g}^{2}+P_{g}$ across the layer and integration of the mass conservation
equation gives continuity of $\rho_{g}u_{g}$ across the layer. This yields
Rankine-Hugoniot equations of the form%
\begin{equation}
\bar{n}_{g}(r_{edge})\ \left(  1+\left(  \bar{u}_{g}(r_{edge})\right)
^{2}\right)  =\ \bar{n}_{g}^{ISM}\left(  \ T_{g}^{ISM}/T_{g}+\ \left(  \bar
{u}_{g}^{ISM}\right)  ^{2}\right)  \label{shock pressure}%
\end{equation}%
\begin{equation}
\bar{n}_{g}(r_{edge})\bar{u}_{g}(r_{edge})=\bar{n}_{g}^{ISM}\ \bar{u}%
_{g}^{ISM} \label{shock continuity}%
\end{equation}
where $\bar{n}_{g}^{ISM},T_{ISM},$ and $T_{g}$ are specified and $\bar{u}%
_{g}^{ISM},r_{edge},$ $\bar{u}_{g}(r_{edge}),$ and $\bar{n}_{g}(r_{edge})$ are
to be determined. The bar indicates that densities are normalized to
$n_{gm}\ $and velocities are normalized to $c_{g}=\sqrt{\kappa T_{g}/m_{g}}.$
Using Eq.\ref{shock continuity} to substitute for $\ \bar{n}_{g}(r_{edge}%
)\bar{u}_{g}(r_{edge})$ in Eq.\ref{shock pressure} gives
\begin{equation}
\bar{n}_{g}(r_{edge})+\left(  \frac{1}{\bar{n}_{g}(r_{edge})}-\frac{1}{\bar
{n}_{g}^{ISM}}\right)  \left(  \bar{n}_{g}^{ISM}\ \bar{u}_{g}^{ISM}\right)
^{2}=\ \ \ \bar{n}_{g}^{ISM}\ \frac{T_{g}^{ISM}}{T_{g}}. \label{elim ug edge}%
\end{equation}
However using Eq.\ref{Mdot gas} evaluated in the ISM at $r_{edge}$,
Eq.\ref{solve Mdot} and Eq.\ref{Jeans radius} give
\begin{equation}
\bar{n}_{g}^{ISM}\bar{u}_{g}^{ISM}=\ -\eta\varepsilon_{BJ}^{2}\ \frac
{r_{J}^{2}}{r_{edge}^{2}}\ \label{solve ugISM}%
\end{equation}
while Eq.\ref{Pade solution} gives
\begin{equation}
\bar{n}_{g}(r_{edge})\simeq\frac{2r_{J}^{2}\ }{\ r_{edge}^{2}\ }%
\ \label{ng(r_edge)}%
\end{equation}
so eliminating $r_{J}^{2}/r_{edge}^{2}$ between these equations gives%
\begin{equation}
\bar{n}_{g}^{ISM}\bar{u}_{g}^{ISM}\simeq-\frac{\eta\varepsilon_{BJ}^{2}}%
{2}\bar{n}_{g}(r_{edge}). \label{comb}%
\end{equation}
Substituting for $\bar{n}_{g}^{ISM}\bar{u}_{g}^{ISM}\ $ in
Eq.\ref{elim ug edge} gives
\begin{equation}
\ -\frac{\left(  \bar{n}_{g}(r_{edge})\right)  ^{2}}{\bar{n}_{g}^{ISM}}\left(
\frac{\eta\varepsilon_{BJ}^{2}}{2}\right)  ^{2}+\bar{n}_{g}(r_{edge})\left(
1+\ \left(  \frac{\eta\varepsilon_{BJ}^{2}}{2}\right)  ^{2}\right)
-\ \ \ \bar{n}_{g}^{ISM}\ \frac{T_{g}^{ISM}}{T_{g}}=0, \label{quadratic shock}%
\end{equation}
a quadratic equation in $\bar{n}_{g}(r_{edge}).$ Since $\varepsilon_{BJ}$ \ is
assumed small, an approximate solution to Eq.\ref{quadratic shock} may be
obtained by balancing the last two terms in which case%
\begin{equation}
\bar{n}_{g}(r_{edge})\simeq\frac{\bar{n}_{g}^{ISM}}{\left(  1+\ \eta
^{2}\varepsilon_{BJ}^{4}/4\right)  }\ \frac{T_{g}^{ISM}}{T_{g}}\ .
\label{edge solution}%
\end{equation}
Equation \ref{ng(r_edge)} then gives%
\begin{equation}
\ \frac{2r_{J}^{2}\ }{\ r_{edge}^{2}\ }=\frac{\bar{n}_{g}^{ISM}}{\left(
1+\ \eta^{2}\varepsilon_{BJ}^{4}/4\right)  }\ \frac{T_{g}^{ISM}}{T_{g}}
\label{elim nedge}%
\end{equation}
so$\ $ using Eq.\ref{Jeans radius}
\begin{equation}
r_{edge}\simeq c_{g}^{2}\sqrt{\frac{1+\ \varepsilon_{BJ}^{4}\eta^{2}/4}{2\pi
Gn_{g}^{ISM}\kappa T_{g}^{ISM}}\ }. \label{redge 2}%
\end{equation}

Equation \ref{solve ugISM} gives%
\begin{equation}
u_{g}^{ISM}=\ -\eta\varepsilon_{BJ}^{2}\ \frac{T_{g}^{ISM}}{2T_{g}}\frac
{c_{g}}{\left(  1+\varepsilon_{BJ}^{4}\eta^{2}/4\right)  },
\label{Ug_ISM solve}%
\end{equation}
which combined with Eq.\ref{shock continuity} and Eq.\ref{edge solution} give
\begin{equation}
u_{g}(r_{edge})=-\ \frac{\varepsilon_{BJ\ }^{2}\eta c_{g}}{2}\ \ .
\label{solve inner velocity at edge}%
\end{equation}

The gas mass between $r_{J}$ and $r_{edge}$ is%
\begin{equation}
M_{r<r_{edge}}\simeq\ \int_{r_{J}}^{r_{edge}}\left(  \rho_{gm}2\frac{r_{J}%
^{2}}{r^{2}}\right)  4\pi r^{2}dr\simeq\ 2c_{g}^{4}\sqrt{\frac{1+\ \varepsilon
_{BJ}^{4}\eta^{2}/4}{2\pi G^{3}n_{g}^{ISM}\kappa T_{g}^{ISM}}\ }.
\label{total gas mass}%
\end{equation}
$\ $ $\ $

The mass inside the core region of radius $r_{J}$ is using
Eqs.\ref{Pade solution}, \ref{Jeans radius} and \ref{density limit}%
\begin{equation}
M_{r<r_{J}}\simeq\ \rho_{gm}\int_{r_{gm}}^{r_{J}}\frac{4\pi r^{2}}%
{1+r^{2}/6r_{J}^{2}}dr\simeq\ \rho_{gm}\frac{4\pi r_{J}^{3}}{3}\simeq
\frac{M_{\ast}}{6\varepsilon_{BJ}}. \label{inner core mass 2}%
\end{equation}

Table \ref{gas solutions} shows quantities predicted by this analysis for
$\varepsilon_{BJ}=0.3$, $\ M_{\ast}=0.4M_{\odot},$ $T_{g}=10$ K, $T_{g}%
^{ISM}=100$ K, molecular hydrogen, and $n_{g}^{ISM}=10^{7}\ $m$^{-3}.$ Note
that the gas flow velocities are subsonic (i.e. much smaller than the random
or thermal velocity) both in the portion of the cloud external to $r_{B}$ and
in the ISM since in the portion of the cloud external to $r_{B}$ gas flow
velocities $u_{g}$ are much slower than $c_{g}=\sqrt{\kappa T_{g}/m_{g}}=200$
m/s and in the ISM the gas flow velocity \ $u_{g}^{ISM}$ is much slower than
$c_{g}^{ISM}=\sqrt{\kappa T_{g}^{ISM}/m_{g}}=640$ m/s.%

\begin{table}[tbp] \centering
\begin{tabular}
[c]{llllll}\hline\hline
& value & units & value & units & Reference\\\hline\hline
$r_{B}$ & $6.5\times10^{14}$ & m & $4.3\times10^{3}$ & au & Eq.\ref{rB}\\
$r_{gm}$ & $1.2\times10^{15}$ & m & $7.9\times10^{3}$ & au & Eq.\ref{def rgm}%
\\
$r_{J}$ & $2.2\times10^{15}$ & m & $1.4\times10^{4}$ & au &
Eq.\ref{Jeans radius}\\
$n_{gm}$ & $3.1\times10^{9}$ & m$^{-3}$ & $3.1\ \times10^{3}$ & cm$^{-3}$ &
Eq.\ref{density limit}\\
$u_{g}(r_{gm})$ & $-91$ & m/s &  &  & Eq.\ref{ug at rgm}\\
$c_{g}$ & $2.0\times10^{2}$ & m/s &  &  & Eq.\ref{cg}\\
$\dot{M}_{g}$ & $1.7\times10^{16}$ & kg /s & $2.7\times10^{-7}$ & $M_{\odot}$
yr$^{-1}$ & Eq.\ref{solve Mdot}\\
$r_{edge}$ & $1.7\times10^{16}$ & m & $1.1\times10^{5}$ & au &
Eq.\ref{redge 2}\\
$r_{edge}/r_{Jeans}$ & $7.7$ &  &  &  & \\
$u_{g}(r_{edge})$ & $-14$ & m/s & $\ $ &  &
Eq.\ref{solve inner velocity at edge}\\
$u_{g}^{ISM}$ & $-1.4\times10^{2}$ & m/s &  &  & Eq.\ref{Ug_ISM solve}\\
$c_{g}^{ISM}$ & $6.4\times10^{2}$ & m/s &  &  & Eq.\ref{cg}\\
$M_{r<r_{edge}}$ & $2.1\times10^{31}$ & kg & $10$ & $M_{\odot}$ &
Eq.\ref{total gas mass}\\
$M_{r<r_{J}}$ & $4.4\times10^{29}$ & kg & $0.2$ & $M_{\odot}$ &
Eq.\ref{inner core mass 2}\\
$M_{r<r_{B}}$ & $1.8\times10^{28}$ & kg & $0.01$ & $M_{\odot}$ &
Eq.\ref{Mass enclosed by rB}\\\hline\hline
\end{tabular}
\caption{Calculated quantities for $ \epsilon_{BJ}$ =0.3, $M_{*}=0.4M_{\odot}$,  $T_g$=10 K, $c_g= 2 \times 10^{2}$ m/s}\label{gas solutions}%
\end{table}%

\section{Dust}

\subsection{Overview of dust}

The mass density of dust in the ISM is well established to be 1\% of the \ gas
mass density
\citep{Lilley1955,Zubko2004}%
. As mentioned earlier, it is commonly assumed that this 1\% ratio also holds
in a dense molecular cloud. However,
\citet{Padoan2006}
have recently cast doubt on such an assumption arguing that\ if the
dust-to-gas mass ratio were indeed 1\% in a dense molecular cloud, then the
dust and column densities should have the same spectral power law, i.e., the
spatial Fourier power spectra $S(k)$ of dust and gas images should be
characterized by the same power law $S(k)\sim k^{-p}$. \ In fact,
\citet{Padoan2006}
found that $p$ for the dust differs from $p$ for the gas, indicating a lack of
local proportionality between dust and gas.%

\citet{Takeuchi2005}
have also pointed out that the dust/gas mass ratio in a cloud need not be the
same as in the ISM because gas and dust might be subject to different
processes.
\citet{Goldsmith1997}
showed in their Fig.22 that the observed dust-to-gas ratio varies by an order
of magnitude as a function of position in a system of molecular clouds, i.e.,
the observed dust-to-gas mass ratio in a molecular cloud system is not, as in
the ISM, fixed at 1\%. As mentioned earlier,
\citet{Fukue2001}
presented numerical calculations specifically demonstrating that accretion can
enrich the dust-to-gas ratio. As also mentioned earlier,
\citet{Fukue2001}
assumed that a Bondi accretion region was directly bounded by the ISM, i.e.,
did not take into account the molecular cloud scale where gas self-gravity is
important. This omission resulted in calculated mass accretion rates lower
than typical observed values (e.g., Fukue's nominal accretion rate was
$\dot{M}_{g}\sim10^{-10}$ $M_{\odot}$ yr$^{-1}$ compared to the $\ $observed
nominal $\dot{M}_{g}\sim10^{-9}$ $M_{\odot}$ to $10^{-7}$ $M_{\odot}$ rates
for pre-main sequence YSO's reported by
\citet{Hartmann1998}%
). Since gas and dust density scale linearly with accretion rate, the low
nominal accretion rate in Fukue's calculation corresponded to a very low dust
density and hence optically thin dust. Higher accretion rates give higher dust
densities and optically thick dust densities that correspond to the observed
opacities of clouds. Nevertheless,\ Fukue's calculation had the interesting
feature of predicting that in certain circumstances the accretion process
could cause a thirty-fold increase in the dust-to-gas mass ratio.

We now argue that this enrichment reported by Fukue should be a basic property
of dust/gas accretion systems, so the dust/gas mass density ratio in a cloud
core should in general \ be substantially enriched relative to its 1\% ISM
ratio. Because the accretion region is now assumed to be bounded by a cloud
core rather than\ by the ISM, this enrichment effect is demonstrated here in
association with more realistic mass accretion rates.\ 

Collision properties of gas and dust vary considerably in ISM, cloud, cloud
core, and accretion regions. This variation is because the scale lengths are
different and because the mean free paths are different. Dust is collisionally
decoupled from gas in some regions, but strongly coupled in others. In
particular, it turns out that dust is decoupled from gas in the ISM, strongly
coupled in the molecular cloud and cloud core, but then can become decoupled
again \ when dust-dust collisions result in dust coagulation. Dust-dust
collisionality is closely related to dust opacity since the mean free path for
dust-dust collisions is $l_{dd}=1/n_{d}\sigma_{d}$ while the distance for
unity optical depth is the same if light scattering is geometric, but larger
by the factor $1/Q_{ext}$, where the efficiency factor $Q_{ext}$ accounts for
weaker than geometric scattering (i.e., Rayleigh) and for absorption.

Let us start with the ISM. Because of the low gas density in the ISM, dust is
collisionally decoupled from gas so if the dust becomes charged,
electromagnetic mechanisms can accelerate ISM\ dust grains to velocities much
higher than the $\sim$ 640 m/s \ ISM gas thermal velocity.
\citet{Meyer1998}%
,
\citet{Yan2003}%
, \
\citet{Slavin2004}%
, and
\citet{Shukla2005}
have provided examples of such dust acceleration mechanisms.

In contrast, dust is collisionally coupled to gas in molecular clouds because
of the high gas density. In cloud cores and/or accretion regions the dust
density becomes so high that dust-dust collisions also become important. The
condition for dust-dust collisions to be important corresponds approximately
to the condition that the dust is optically thick. Thus, if there is an inner
region where dust-dust collisions are important, regions external to this
inner region will be shielded from central object optical emission and so will
not experience any radiation pressure from the central object. Dust-dust
collisions result in dust coagulation, a condition where the dust radius
$r_{d}$ increases while the overall dust mass density remains invariant.
Coagulation has the dual effect of causing the dust to revert to being
optically thin and collisionless. Collisionless, coagulated dust exposed to
stellar radiation becomes charged via photo-emission and so must be described
by dusty plasma dynamics rather than by gas dynamics. This dusty plasma regime
will be considered in a future publication; preliminary results are presented
in
\citet{Bellan2007}%
.

\subsection{Dust grain size assumption}

Mathis, Rumpl, and Nordsieck (1977) reported observations showing that ISM
\ dust grains have a size distribution scaling as $r_{d}^{-3.5}$ with a
nominal lower limit radius $r_{low}=0.005$ $\mu$m and a nominal upper limit
$r_{high}=0.25$ $\mu$m. This size distribution is commonly referred to as the
MRN dust size distribution.
\citet{Miyake1993}
proposed that the dust size distribution in a circumstellar disk will have
larger grains than given by the MRN\ distribution because in a circumstellar
disk where the dust density is high, dust coagulates due to dust-dust
collisions. More recently,
\citet{Przygodda2003}
and
\citet{vanBoekel2003}
have reported direct observational evidence of grain growth in circumstellar
disks while, in addition,
\citet{Dullemond2005}
provided detailed calculations showing a strong tendency for dust grain growth
when dust grains collide with each other.

The $r_{d}^{-3.5}$ dust size distribution has important well-known statistical
properties. Specifically, an $r_{d}^{-3.5}\ $distribution implies (i) most of
the mass is contained in the very small number of heavy grains and (ii)\ most
of the surface area is contained in the very large number of light grains
\citep{Smith1998}%
. Although the lightest dust grains provide most of the surface area, their
radii are so much smaller than visible light wavelengths that their scattering
is Rayleigh rather than geometric. Rayleigh scattering is weaker than
geometric scattering by a factor $r_{d}^{4}/\lambda^{4}$ where $\lambda$ is
the wavelength of the radiation. Hence, despite being numerous, the very small
radius grains are ineffectual at contributing to the optical depth at visible
wavelengths so visible scattering will be due mainly to the larger grains
which also contain most of the mass. Hence, for purposes of both mass
inventory and optical depth we will consider that dust grains have a nominal
radius $r_{d}=$ $0.1$ $\mu$m.

\subsection{Dust velocity in the ISM}

If the velocity of individual dust grains in the ISM were the result of
thermal equilibration with gas molecules, the dust kinetic temperature would
be the same as the gas molecule kinetic temperature in which case the random
velocity of dust grains in the ISM\ would be $\ u_{d}^{ISM}\sim\sqrt{\kappa
T_{g}/m_{d}}=\ \ 0.01$ m/s. This velocity is so small that there would be
insufficient dust flux entering a molecular cloud to populate the cloud with
dust in a reasonable time. It has been proposed by several authors that dust
attains a much larger random velocity in the ISM because dust is nearly
collisionless in the ISM so that collective collisionless mechanisms such as
shocks or turbulence could accelerate dust to very large velocities. As
mentioned earlier,
\citet{Meyer1998}%
,
\citet{Yan2003}%
,
\citet{Slavin2004}%
, and
\citet{Shukla2005}
have presented collisionless mechanisms whereby dust grains are accelerated to
velocities in the 10 km/s range.

A typical argument for collisionless acceleration is that a spatially and
temporally random magnetic field develops with energy density in thermodynamic
equipartition with the gas, i.e., $B^{2}/2\mu_{0}\sim n_{g}^{ISM}\kappa
T_{g}^{ISM}$ so
\begin{equation}
B\sim\sqrt{2\mu_{0}n_{g}^{ISM}\kappa T_{g}^{ISM}}\sim2\times10^{-10}%
\quad\mathrm{T.} \label{B_ISM}%
\end{equation}
Gradients in this microgauss magnetic field then accelerate charged dust
grains via electromagnetic forces so that the dust random kinetic energy
becomes comparable to the magnetic energy density, i.e., $\ $%
\begin{equation}
\frac{1}{2}\rho_{d}^{ISM}\left(  U_{d}^{ISM}\right)  ^{2}\sim\frac{B^{2}}%
{2\mu_{0}}\ \ \label{dust equipartion with B}%
\end{equation}
where $U_{d}^{ISM}$ is a random velocity that could be pointing in any
direction. $\ $The nominal random velocity $U_{d}^{ISM}$ attained by the
accelerated grains is thus the dust Alfv\'{e}n velocity given by $v_{A}%
^{2}=B^{2}/\mu_{0}\rho_{d}^{ISM}$. Combining Eqs.\ref{B_ISM} and
\ref{dust equipartion with B} gives the random velocity to be
\begin{equation}
U_{d}^{ISM}\simeq\ \sqrt{\frac{2\kappa T_{g}^{ISM}}{m_{g}}\frac{\rho_{g}%
^{ISM}}{\rho_{d}^{ISM}}}, \label{UdISM}%
\end{equation}
a relationship dependent on the dust-to-gas mass density ratio. Equation
\ref{UdISM} gives $U_{d}^{ISM}\sim9.3$ km/s for a nominal 1\% ISM\ dust-to-gas
mass ratio and $T_{g}^{ISM}=100$ K. This velocity is six orders of magnitude
larger than the velocity predicted by thermal equilibration of individual dust
grains with gas molecules. Because the dust is assumed to be accelerated by
collisionless processes, it will not in general have a Maxwellian
distribution, and in fact it would be reasonable to expect that each dust
grain has a velocity $\sim U_{d}^{ISM}$ but with a random direction. This
situation would be similar to the neutrons emitted from a fission reaction
since such neutrons all have the same kinetic energy but have random velocity directions.

The mean radial velocity of dust grains entering a spherical molecular cloud
can be determined by considering the radial velocity component of these dust
grains at the surface of the sphere. If $\hat{s}$ denotes the inwards normal
to the surface at some point on the surface of the sphere, then the radial
velocity of a dust grain entering the sphere at this point is $U_{d}^{ISM}%
\cos\theta$ where $\theta\,\ $\ is the angle between the dust grain vector
velocity and $\hat{s}.$ The angle $\theta$ must lie between $-\pi/2$ and
$\pi/2$ since dust grains having $\theta$ outside this range will not enter
the sphere. Other than being restricted to the range $-\pi/2$ $<\theta<$
$\pi/2,$ the angle $\theta$ is random. The dust grains have equal probability
of having any angle $\theta$ between $0$ and $2\pi$ so the probability of a
dust grain angle being between $\theta$ and $\theta+d\theta$ is $d\theta
/2\pi.$ The mean radial velocity of dust grains entering a molecular cloud
from the ISM is thus
\begin{equation}
u_{d}^{ISM}=\ \int_{-\pi/2}^{\pi/2}U_{d}^{ISM}\cos\theta\,\frac{d\theta}{2\pi
}=\frac{1}{\pi}U_{d}^{ISM}\simeq3\mathrm{\ km/s.} \label{calculate udISM}%
\end{equation}
The question of whether dust exits a molecular cloud will be addressed in the
next section.

\subsection{Dust entrainment by gas in molecular cloud}

The dust equation of motion can be expressed as%
\begin{equation}
\ u_{d}\frac{du_{d}}{dr}\mathbf{\ }=\ \ -\ \frac{G}{r^{2}}\left(  M_{\ast
}+\int_{r_{\ast}}^{r}4\pi\xi^{2}\left(  \rho_{g}(\xi)+\rho_{d}(\xi)\right)
d\xi\right)  \ -\ (u_{d}-u_{g})\frac{\rho_{g}}{m_{d}}\sigma_{d}\sqrt{c_{g}%
^{2}+(u_{d}-u_{g})^{2}} \label{dust motion 2}%
\end{equation}
where it is recalled that both $u_{d}$ and $u_{g}$ are negative corresponding
to radial inward motion.

We wish to investigate the possibility that $\rho_{d}/\rho_{g}$ could increase
from its 1\% ISM\ value as dust and gas enter the cloud, core, and Bondi
accretion region. However, we will assume that $\rho_{d}/\rho_{g}$ $\ $remains
sufficiently small compared to unity that the back-reaction of dust on gas
dynamics can be ignored. Thus, $u_{g}(r)$ and $\rho_{g}(r)$ are assumed to
remain as discussed in Sec.\ref{Bondi regime} and \ref{Jeans regime}. Dust
enters the cloud from the ISM with a velocity much higher than the gas thermal
and fluid velocities but then collides with the gas, thereby slowing down
until moving at nearly the same inward fluid velocity as the gas. Collisions
thus cause the dust to become entrained by the gas inflow
\citep{Fukue2001}%
.

The collisional slowing down can be estimated by assuming that the
gravitational force term in Eq.\ref{dust motion 2} can be ignored because of
the high dust velocity. Furthermore, $u_{d}\ $is much larger than both
$u_{g}\ $and $c_{g}$ when the dust enters the cloud from the ISM. In this case
and using%
\begin{equation}
\frac{\sigma_{d}}{m_{d}}=\frac{3}{4r_{d}\rho_{d}^{int}} \label{sigma md ratio}%
\end{equation}
where $\rho_{d}^{int}\simeq2\times10^{3}$ kg m$^{-3}$ is the intrinsic mass
density of the dust, Eq.\ref{dust motion 2} becomes%
\begin{equation}
\frac{du_{d}}{dr}\mathbf{\ }=-\ u_{d}\ \frac{3\rho_{g}(r)}{4r_{d}\rho
_{d}^{int}}. \label{simple slowing}%
\end{equation}
Integration of Eq. \ref{simple slowing} starting from the cloud edge and going
inwards gives
\begin{equation}
u_{d}(r)=u_{d}(r_{edge})\exp\left(  -\int_{r}^{r_{edge}}\ \frac{3\rho_{g}%
(\xi)}{4r_{d}\rho_{d}^{int}}d\xi\right)  . \label{solve slowing}%
\end{equation}
The dashed line in Fig. \ref{f1.eps}(b) shows the dust velocity given by
Eq.\ref{solve slowing} for the gas density given in Fig. \ref{f1.eps}(a) and
assuming a 3 km/s dust mean radial entrance velocity. It is seen that the dust
rapidly slows down until it achieves the same velocity as the gas at which
point it is assumed to be entrained by the gas. Dust will thus become
entrained by gas at a radius $r_{entrain}$ where $r_{entrain}$ satisfies
\begin{equation}
u_{d}^{ISM}\exp\left(  -\int_{r_{entrain}}^{r_{edge}}\ \frac{3\rho_{g}%
(r)}{4r_{d}\rho_{d}^{int}}dr\right)  \sim u_{g}^{edge}\sim c_{g}/20.
\label{entrainment}%
\end{equation}
Using Eq.\ref{calculate udISM} to estimate $u_{d}^{ISM}$, Eq.\ref{entrainment}
becomes
\begin{equation}
\int_{r_{entrain}}^{r_{edge}}\ \frac{3\rho_{g}(r)}{4r_{d}\rho_{d}^{int}}%
dr\sim\ln\left(  \frac{20u_{d}^{ISM}}{c_{g}}\right)  \simeq5.7\qquad.
\label{stopping 1}%
\end{equation}
If the gas density in the $r>>r_{J}$ portion of the cloud is sufficiently
large to provide entrainment, then $\rho_{g}(r)$ can be approximated in this
region as $\rho_{g}(r)\simeq2\rho_{gm}r_{J}^{2}/r^{2}$ so Eq.\ref{stopping 1}
becomes
\begin{equation}
\frac{3\rho_{gm}r_{J}}{2r_{d}\rho_{d}^{int}\ }\left(  \frac{r_{J}}%
{r_{entrain}}-\frac{r_{J}}{r_{edge}}\right)  \simeq5.7\qquad\label{stopping 3}%
\end{equation}
Table 3 gives $r_{edge}/r_{J}\simeq7.7$ and since $3\rho_{gm}r_{J}/2\rho
_{d}^{int}r_{d}=1.7\times10^{2}$ this gives $r_{entrain}\simeq6r_{J}.$ As
assumed, the dust becomes entrained by the gas at a location in the $r>>r_{J}$
portion of the cloud.

The entrainment of dust by gas means that the radial flow of dust from the
ISM\ to the cloud is one-way inwards, i.e., there is no radial outward flow of
individual dust grains from the cloud back to the ISM. This one-way inward
behavior of dust grains is in contrast to gas. Because gas molecules \ collide
with each other, some gas molecules have inwards radial velocities and some
have outwards radial velocities. However, there are more inwards than outwards
moving gas molecules and the net inwards gas fluid velocity at $r_{edge}$ is a
consequence of this difference. The situation is somewhat analogous to a home
vacuum cleaner ingesting macroscopic particles and air; individual air
molecules in the vacuum hose have inwards or outwards velocities with
magnitude of the order of the air thermal velocity, the mean air velocity in
the hose is inwards and much slower than the air thermal velocity, and all the
macroscopic particles in the hose have inwards velocities.

\subsection{Enrichment of dust-to-gas mass ratio\label{dust enrichment}}

Equations \ref{Mdot gas} and \ref{Mdot dust} can be evaluated at any radius.
Evaluating at $r_{edge}$ and at $r_{gm}$ gives%
\begin{equation}
4\pi r_{edge}^{2}\rho_{g}^{ISM}u_{g}^{ISM}=4\pi r_{gm}^{2}\rho_{g}^{\ }%
(r_{gm})u_{g}^{\ }(r_{gm}) \label{gas flux balance}%
\end{equation}%
\begin{equation}
4\pi r_{edge}^{2}\rho_{d}^{ISM}u_{d}^{ISM}=4\pi r_{gm}^{2}\rho_{d}^{\ }%
(r_{gm})u_{d}^{\ }(r_{gm}). \label{dust flux balance}%
\end{equation}

Dividing Eq.\ref{gas flux balance} by Eq.\ref{dust flux balance} gives%
\begin{equation}
\frac{\rho_{d}^{\ }(r_{gm})/\rho_{g}^{\ }(r_{gm})\ }{\rho_{d}^{ISM}/\rho
_{g}^{ISM}}=\ \frac{u_{d}^{ISM}/u_{g}^{ISM}}{u_{d}(r_{gm})/u_{g}(r_{gm}%
)\ }\ \ . \label{enrichment}%
\end{equation}

Since the dust is certainly fully entrained by the time it reaches $r_{gm}$ we
may assume $u_{d}(r_{gm})/u_{g}(r_{gm})=1.$ Using Eq.\ref{Ug_ISM solve} to
give $u_{g}^{ISM}$ and Eqs. \ref{UdISM} and \ref{calculate udISM} to give
$u_{d}^{ISM}$, Eq.\ref{enrichment} becomes
\begin{equation}
\frac{\rho_{d}^{\ }(r_{gm})}{\ \rho_{g}^{\ }(r_{gm})\ }=\frac{2+\varepsilon
_{BJ}^{4}\eta^{2}/2}{\pi\eta\varepsilon_{BJ}^{2}\ \ \ }\sqrt{\ \frac{2T_{g}%
}{T_{g}^{ISM}}\frac{\rho_{d}^{ISM}}{\rho_{g}^{ISM}}}.\label{enrichment 2}%
\end{equation}
Fig. \ref{f1.eps}(c)\ shows the dust density and Fig.\ref{f1.eps}(d)\ shows
the dust-to-gas mass ratio as function of radius, showing that enrichment
occurs so that the dust-to-gas ratio increases from its 0.01 ISM\ value to
become somewhat less than unity. For the example parameters in Table 3 and a
1\% ISM dust-to-gas mass ratio, Eq.\ref{enrichment 2} predicts $\rho_{d}%
^{\ }(r_{gm})/\rho_{g}^{\ }(r_{gm})=\ \allowbreak0.2\,$ for $\varepsilon
_{BJ}=0.3$, $T_{g}=10$ K, and $T^{ISM}$ $=100$ K; this corresponds to the
saturated value in Fig.\ref{f1.eps}(d) and represents a twenty-fold enrichment
of the dust-to-gas mass ratio over its ISM\ value. %

\begin{figure}
[h]
\begin{center}
\includegraphics[
height=1.7616in,
width=5.7804in
]%
{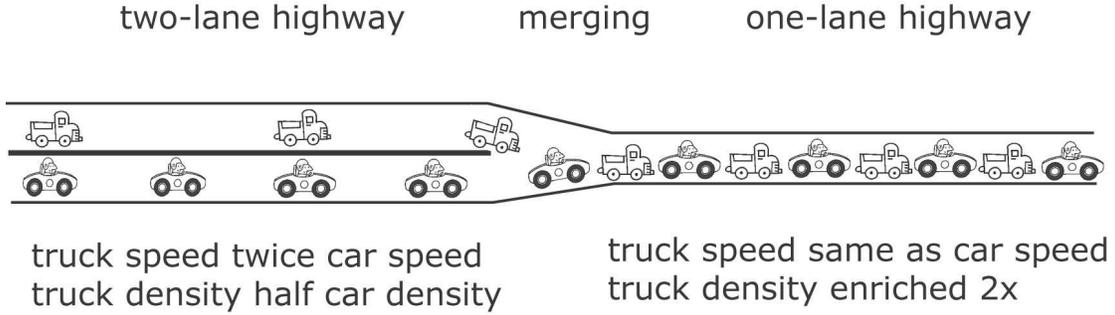}%
\caption{Because trucks\ entering the one-lane segment slow down whereas cars
do not, the truck-to-car density ratio is enriched in the one-lane segment
compared to the density ratio in the two-lane segment.}%
\label{CAR TRUCK ENRICHMENT Jan-15-2008-VerdanaFont.eps}%
\end{center}
\end{figure}

An everyday example of the enrichment effect predicted by Eq.\ref{enrichment}
is where traffic on a two-lane highway merges to flow on a one-line highway as
sketched in Fig.\ref{CAR TRUCK ENRICHMENT Jan-15-2008-VerdanaFont.eps}. In the
two-lane segment (left of figure), the trucks move at twice the speed of the
cars and have half the density of the cars. The cars are assumed to have the
same speed in the two-lane and one-lane highway segments, i.e., the cars do
not slow down. The trucks have to slow down when they enter the one-lane
segment because in the one-lane segment the trucks must go at the same speed
as the cars. Slowing down by a factor of two causes the number of trucks per
length of road to double because in steady-state the flux of trucks (product
of the truck speed and the trucks per length of road) must be the same in the
one- and two-lane segments. The density of trucks to cars is thus enriched by
a factor of two in the one-lane highway segment compared to its value in the
two-lane highway segment.

\subsection{Radiation pressure and optical depth}%

\citet{Abbas2003}
used a calibrated electrodynamic balance to measure the radiation pressure
force exerted on a $r_{d}=$0.125 $\mu$m \ particle by a 0.532 nm wavelength
laser. They found that the force exerted by a laser intensity $8\times10^{5}$
W m$^{-2}$ was $3\times10^{-17}$ N. If such a particle were in the
gravitational field of a star with luminosity $L_{\ast}$, the radiation
pressure force on the particle would thus be
\begin{equation}
F_{rad}=\frac{L_{\ast}e^{-\tau(r)}}{4\pi r^{2}}\frac{3\times10^{-17}}%
{8\times10^{5}}\quad\mathrm{N} \label{Frad}%
\end{equation}
where%
\begin{equation}
\tau(r)=\int_{r_{\ast}}^{r}Q_{ext}n_{d}(\xi)\sigma_{d}d\xi\label{tau}%
\end{equation}
is the optical depth for light traveling from the star to radius $r\ $ and
$Q_{ext}$ is the extinction efficiency. The gravitational force would be%
\begin{equation}
F_{g}=\frac{m_{g}M_{\ast}G}{r^{2}}\quad\mathrm{N} \label{Fgrav}%
\end{equation}
so the ratio would be%
\begin{equation}
\frac{F_{rad}}{F_{g}}=\frac{L_{\ast}e^{-\tau(r)}}{4\pi\ }\frac{3\times
10^{-17}}{8\times10^{5}}\frac{\ 1}{m_{g}M_{\ast}G}. \label{F ratio}%
\end{equation}
Assuming $m_{g}=10^{-17}$ kg for a nominal $r_{d}=10^{-7}$ m dust grain gives
\begin{equation}
\frac{F_{rad}}{F_{g}}=\ 0.9e^{-\tau(r)}\frac{L_{\ast}/L_{\odot}}{M_{\ast
}/M_{\odot}}\ \label{Fratio norm}%
\end{equation}
so a dust grain could in principle be subject to significant radiation force
if $\tau\ll1\ $as was proposed by Fukue. Table 1 of
\citet{Abbas2003}
gives the calculated radiation pressure efficiency $Q_{pr}=0.28\ $for a
\ $r_{d}=0.125$ $\mu$m dust grain illuminated by 532 nm light. This is in
reasonable agreement with their measured value\ $Q_{pr}=0.22;$
\citet{Abbas2003}
also presented a calculated extinction efficiency $Q_{ext}=0.33$ and a
calculated scattering efficiency $Q_{sca}=0.16$ that together would be in
reasonable agreement with the measured $Q_{pr}$. Figure \ref{f1.eps}(c) shows
that dust for the representative parameters discussed here has a density
$n_{d}\simeq1$ m$^{-3}$ at $r_{B}\ $ and that this density increases for
$r<r_{B}.$ Table 3 gives $r_{B}=6.5\times10^{14}$ m. Thus, a lower bound for
the optical depth at $r_{B}$ can be estimated using
\begin{equation}
\tau(r_{B})>\ Q_{ext}n_{d}(r_{B})\sigma_{d}r_{B}=0.33\times1\times\pi
\times(10^{-7})^{2}\times6.5\times10^{14}=6.7\quad. \label{attenuation}%
\end{equation}
Star light would thus be attenuated by a factor greater than $\exp
(-6.7)\simeq10^{-3}$ and so optical radiation from the star should be fully
extinguished by the time it reaches $r_{B}$ in which case there would be no
significant radiation pressure on dust\ at radii of the order of $r_{B}$ or
larger. The reasons why we differ from Fukue by concluding that radiation
pressure is unimportant are (i) Fukue's adiabatic assumption resulted in gas
that was 1600 times hotter at small $r$ so for a given pressure the gas
density would be 1600 times less and so for a given dust-to-gas ratio, the
dust density would be much lower, (ii) Fukue also used a much lower value of
$Q_{ext}\ $than the value given in
\citet{Abbas2003}%
.

\section{Discussion}

The logical requirement that the Bondi length must be smaller than the Jeans
length introduces the dimensionless parameter $\varepsilon_{BJ}$ given by the
ratio of these lengths. An intermediate distance $r_{gm}$ can be defined as
the geometric mean of the Bondi and Jeans lengths so, if the two lengths are
well separated, $r_{gm}$ is much larger than the Bondi length but much smaller
than the Jeans length. Thus, $r_{gm}$ constitutes the large $r$ limit of the
solution to the Bondi problem and the small $r$ limit of the solution to the
Jeans problem. Matching the Bondi and Jeans solutions at $r_{gm}$ provides a
solution valid over scales ranging from smaller than the Bondi length to
larger than the Jeans length. This allows accounting for transonic flow, mass
accretion, and gravitational self-confinement in one self-consistent solution
to the gas equation.

By assuming that the dust mass density and energy density are not larger than
the corresponding gas densities, the gas equations can be evaluated ignoring
interaction with dust. Once the gas behavior has been worked out, the dust can
be considered as moving through a pre-determined gas profile that retards the
dust due to frictional drag. \ The dust slows down greatly whereas the gas
velocity changes relatively little, a distinction that enriches the
dust-to-gas mass ratio. Plausible values of $\varepsilon_{BJ}$ suggest that
the dust-to-gas mass ratio will always be enriched to be somewhat less than
unity, because at unity the assumption that the gas is unaffected by dust
fails. \ The high dust density at small radius means that dust-dust collisions
will become important and cause coagulation of dust. This coagulation will
cause the dust to become optically thin and collisionless again. Because the
coagulated dust is optically thin, it will absorb photons from the star and
become electrically charged due to photo-emission of electrons.
Two-dimensional motion of collisionless, charged dust in a gravitational field
resulting in a poloidal field dynamo has been discussed in
\citet{Bellan2007}%
; three dimensional motion and resulting poloidal/toroidal field dynamo action
will be discussed in a future publication.

The author wishes to thank an anonymous referee for his/her thoughtful comments.

\bigskip
\bibliographystyle{authordate1}
\bibliography{ApJ-gravitydynamoJuly2006}

\end{document}